\documentclass[pra,twocolumn,superscriptaddress,showpacs,preprintnumbers,amsmath,amssymb]{revtex4}
\usepackage{bm}
\usepackage{bbm}
\usepackage{amssymb}
\usepackage{amsfonts}
\usepackage{epsfig,graphicx}
\usepackage{amstext}
\usepackage{amsmath}
\usepackage{graphicx}
\usepackage{times}
\usepackage{dcolumn}
\usepackage{dcolumn}
\usepackage{txfonts}
\usepackage{color}

\begin{document}

\title{Nonlocal advantage of quantum coherence in high-dimensional states}

\author{Ming-Liang Hu}
\email{mingliang0301@163.com}
\affiliation{School of Science, Xi'an University of Posts and Telecommunications, Xi'an 710121, China}
\affiliation{Institute of Physics, Chinese Academy of Sciences, Beijing 100190, China}
\author{Heng Fan}
\email{hfan@iphy.ac.cn}
\affiliation{Institute of Physics, Chinese Academy of Sciences, Beijing 100190, China}
\affiliation{School of Physical Sciences, University of Chinese Academy of Sciences, Beijing 100190, China}
\affiliation{Collaborative Innovation Center of Quantum Matter, Beijing 100190, China}

\begin{abstract}
By local measurements on party \textit{A} of a system \textit{AB}
and classical communication between its two parties, one can achieve
a nonlocal advantage of quantum coherence (NAQC) on party
\textit{B}. For the $l_1$ norm of coherence and the relative entropy
of coherence, we generalized the framework of NAQC for two qubits
and derived the criteria which capture NAQC in the $(d\times
d)$-dimensional states when $d$ is a power of a prime. We also
presented a new framework for formulating NAQC, and showed through
explicit examples its capacity on capturing the NAQC states.
Moreover, we proved that any bipartite state with NAQC is quantum
entangled, thus the obtained criteria can also be used as an
entanglement witness.
\end{abstract}


\maketitle

\section{Introduction} \label{sec:1}
Coherence is a basic notion in quantum theory \cite{Ficek}. It stems
from the superpositions of a set of basis states, and intuitively, a
state is said to be coherent provided that there are nonzero
elements in the nondiagonal position of its density matrix
representation. Coherence is thought to be fundamental and crucial
as it not only can deepen our understanding about the essence of
quantum theory, but also can be used as a physical resource for
developing many fascinating information processing tasks which
outperform their classical counterparts \cite{Plenio,Hu}.

While being widely accepted as a feature unique from classical
physics, it has only recently been suggested to characterize
coherence in a quantitative way. Baumgratz \textit{et al.}
formulated the framework for defining a faithful coherence measure,
and proposed the $l_1$ norm of coherence and relative entropy of
coherence \cite{coher}. Inspired by this, a number of other
coherence measures have also been proposed \cite{meas6,asym1,
co-ski2,meas4,measjpa,new1,dist2}. This sets the stage for a
quantitative study of coherence, with much progress being achieved
in recent years. Some notable ones include their interpretation from
an operational perspective \cite{dist2,distill,qsm}, the cohering
and decohering power of a channel \cite{power1,power2, power3}, the
frozen phenomenon of coherence in noisy environments
\cite{fro1,fro2}, and how coherent states work for improving
efficiency of certain quantum information processing tasks
\cite{meas6,co-ski2,DJ,Grover, DQC1}. There are also works
concentrating on the pivotal role of coherence in capturing the wave
nature of a system \cite{path1,path2} and the maximum coherence in
the optimal basis \cite{mc1,mc2}.

Quantum coherence could also be linked to quantum correlations,
although they were defined in different scenarios (the former is for
the single-partite states, while the latter is for the bipartite and
multipartite states) and capture different aspects of the
quantumness of a system. The essence for this intimate connection
may be their origin of superposition principle, thus it is quite
natural to ask how can one resource be converted to the other
\cite{DQC1}. In fact, inspired by the recent progresses on the
resource theory of coherence, the interplay between coherence and
quantum correlations has attracted people's increasing interest,
e.g., the relations between coherence and entanglement \cite{meas1}
or between coherence and discordlike quantum correlations
\cite{DQC1,Yao,Hufan,Huxy} have already been established.

Apart from the aforementioned progresses, the interplay of coherence
and quantum correlations can also be identified by scrutinizing the
steered local coherence at one part of a bipartite system, while
some works have been accomplished in this direction
\cite{Huxy,steer,NAQC,NAQC2}. In particular, Mondal \textit{et al.}
proposed to think about steerability of the local coherence through
a game between two players, Alice and Bob \cite{NAQC}. By
concentrating on the two-qubit system ${AB}$, they showed that by
local operations on party $A$ and classical communication between
the two parties, the conditional states of $B$ can achieve a
nonlocal advantage of quantum coherence (NAQC). It is then natural
to quest whether a general bipartite state can also achieve the
NAQC. We consider in this paper such a problem. We first generalize
the framework for two qubits to high-dimensional states, then
present a new framework for formulating the NAQC. We will also show
that for any bipartite state that can achieve a NAQC, one is sure
that it is quantum entangled.

\section{Measures of coherence} \label{sec:2}
First, we recall how to quantify coherence in a state. In general,
the starting point for such a quantification is the identification
of incoherent states and incoherent operations. Within the framework
established by Baumgratz \textit{et al.} \cite{coher}, a state is
said to be incoherent if it is diagonal in the given reference basis
$\{|i\rangle\}$, and the incoherent operations are those which map
the incoherent states into incoherent states. This is in direct
analogy to the resource theory of entanglement \cite{ent} and
quantum discord \cite{QD}. Starting from this framework, Baumgratz
{\it et al.} \cite{coher} introduced the defining conditions for a
faithful coherence measure, and proposed to define measures of
coherence as a distance to the closest incoherent states. They also
defined the $l_1$ norm of coherence and relative entropy of
coherence, which are given by \cite{coher}
\begin{equation} \label{eq2-1}
 C_{l_1}(\rho)=\sum_{i\neq j}|\langle i| \rho |j\rangle|,~~
 C_{re}(\rho)=S(\rho_{\mathrm{diag}})-S(\rho),
\end{equation}
where $S(\rho)$ is the von Neumann entropy of $\rho$, and
$\rho_{\mathrm{diag}}$ is the diagonal part of $\rho$. A relation
between $C_{l_1}(\rho)$ and $C_{re}(\rho)$ was established in Ref.
\cite{Rana}. Moreover, $C_{re}(\rho)$ equals to the optimal rate of
the distilled maximally coherent states by incoherent operations in
the asymptotic limit of many copies of $\rho$ \cite{dist2}, and this
endows it with an actual meaning.

Quantum coherence in a state $\rho$ can also be quantified from
other perspectives. In fact, most of the recently proposed measures
were based on the framework of Baumgratz \textit{et al.}
\cite{coher}, with however the different distance measures of states
being adopted. Of course, there are also coherence measures which
were defined by relaxing the defining conditions or by redefining
the free operations. See Refs. \cite{Plenio,Hu} for a review of
these coherence measures.

\section{NAQC in high-dimensional states} \label{sec:3}
For two-qubit states, the criteria for capturing NAQC were
established in Ref. \cite{NAQC}. Here, we derive the criterion for
capturing NAQC in the $(d\times d)$-dimensional states, with $d$
being any power of a prime. Our starting point is the
complementarity relation of coherence under mutually unbiased bases
\cite{comple1}. It states that for single-partite state $\rho$ of
dimension $d$, we have
\begin{equation} \label{eq3-1}
    \sum_{j=1}^{d+1} C^{A_j}(\rho) \leqslant C^m,
\end{equation}
where $C^{A_j}(\rho)$ denotes any faithful coherence measure defined
in the reference basis spanned by the eigenvectors of $A_j$, and
$C^m$ is a state-independent upper bound that cannot be exceed by
any $\rho$. Moreover, $\{A_j\}$ denotes the set of mutually unbiased
observables, and by saying two observables are mutually unbiased, we
mean that the bases comprising their eigenvectors are mutually
unbiased \cite{MUB1,MUB2}.

Now, we extend the framework of the NAQC for two-qubit states
\cite{NAQC} to a more general scenario. Without loss of generality,
we suppose Alice and Bob share a $(d\times d)$-dimensional state
$\rho_{AB}$. Before the game commences, they agree on the set of
measurements $\{A_i\}$. Alice then carries out one of the
measurements chosen at random and informs Bob of her choice $A_i$
and outcomes $a$. Bob's task is to measure the coherence of his
conditional states in the basis spanned by the eigenvectors of all
possible $A_j$ other than $j= i$. After Alice performing all the
possible measurements $\{A_i\}$ with equal probability, Bob's
coherence averaged over all of his possible conditional states and
all of his allowable bases is given by
\begin{equation} \label{eq3-2}
  C^{na}(\rho_{AB})= \frac{1}{d}\sum_{i,j,a \atop i\neq j} p(a|A_i)
                     C^{A_j}(\rho_{B|A_i^a}),
\end{equation}
where $p(a|A_i)$ is the probability for Alice's measurement outcome
$a$ when measuring $A_i$, and $\rho_{B|A_i^a}$ is the conditional
state of Bob (see Appendix \ref{sec:A}).

Based on the protocol stated above, one can establish a general
criterion for capturing NAQC in $\rho_{AB}$. It reads
\begin{equation} \label{eq3-3}
  C^{na}(\rho_{AB})> C^m,
\end{equation}
which is an immediate result of Eq. \eqref{eq3-1} as the bound $C^m$
is not achievable for any single-partite state of dimension $d$. In
the following, we will say $\rho_{AB}$ a NAQC state if it obeys Eq.
\eqref{eq3-3}.

Using the criterion \eqref{eq3-3}, one can prove that there are no
separable states that can achieve the NAQC. This is because for any
$\rho_\mathrm{sep}=\sum_k q_k \rho_A^k \otimes \rho_B^k$, the
conditional state of Bob is given by (see Appendix \ref{sec:A})
\begin{equation} \label{eq3-4}
 \rho_{B|A_i^a}= \frac{\sum_k q_k p_k(a|A_i) \rho_B^k}
                 {p(a|A_i)}.
\end{equation}
where $p_k (a|A_i)=\langle\phi_a^i| \rho_A^k |\phi_a^i\rangle$, and
$p(a|A_i)=\sum_k q_k p_k (a|A_i)$. Then by using the convexity of
the coherence measure $C$, one can obtain
\begin{equation} \label{eq3-5}
 \begin{aligned}
  C^{na}(\rho_\mathrm{sep})&= \frac{1}{d}\sum_{i,j,a \atop i\neq j} p(a|A_i)C^{A_j}(\rho_{B|A_i^a}) \\
                           &\leqslant \frac{1}{d}\sum_{i,j,k,a \atop i\neq j}q_k
                                      p_k(a|A_i) C^{A_j}(\rho_B^k) \\
                           &=\sum_{j,k} q_k C^{A_j}(\rho_B^k)
                           \leqslant C^m,
 \end{aligned}
\end{equation}
where the second equality is due to $\sum_a p_k (a|A_i)=1$ ($\forall
i,k$), and the last inequality is because $\rho_B^k$ may not be the
optimal state for saturating Eq. \eqref{eq3-1}. This completes the
proof.

For a bipartite entangled state $\rho_{AB}$, it is also possible
that $C^{na}(\rho_{AB})\leqslant C^m$, while Eq. \eqref{eq3-5}
implies that all the NAQC states are entangled, hence one may
recognize what the NAQC captures as a kind of quantum correlation in
$\rho_{AB}$ which is stronger than quantum entanglement. Moreover,
as Eq. \eqref{eq3-1} is derived based on the complete set of the
$d+1$ mutually unbiased bases, which are only known to exist when
$d$ is a prime power \cite{MUB2}, our criteria also require $d$ to
be a power of a prime. The identification of a general criterion for
any Hilbert space dimension $d$ is still a fascinating problem of
future research.

Similar to the two-qubit states, the bound $C^m$ obtained with
different coherence measures may be different. The violation of any
one of them by Bob's conditional states implies the existence of
NAQC in $\rho_{AB}$. That is, Eq. \eqref{eq3-3} provides a
sufficient coherence steering criterion. To be explicit, we consider
in the following two faithful coherence measures, i.e., the $l_1$
norm of coherence and the relative entropy of coherence.

\subsection{$l_1$ norm of coherence} \label{sec:3a}
We denote by $C_{l_1}^{A_j}(\rho)$ for the $l_1$ norm of coherence
defined in the basis spanned by eigenvectors of $A_j$. Then for any
single-partite state $\rho$ of dimension $d$, we always have
\cite{comple1}
\begin{equation} \label{eq3a-1}
    C_{l_1}^{A_j}(\rho) \leqslant \sqrt{d(d-1)[P(\rho)-P(A_j|\rho)]},
\end{equation}
where $P(\rho)=\mathrm{tr}\rho^2$, $P(A_j|\rho)=\sum_a \langle
a_j|\rho|a_j\rangle^2$, and $\{|a_j\rangle\}_{a=1}^d$ represent the
eigenvectors of $A_j$. By combining this equation with the mean
inequality (i.e., the arithmetic mean of a list of nonnegative real
numbers is not larger than the quadratic mean of the same list), one
can obtain
\begin{equation} \label{eq3a-2}
  \begin{aligned}
   \sum_{j=1}^{d+1}C_{l_1}^{A_j}(\rho)
     & \leqslant \sqrt{d(d^2-1)\Big[(d+1)P(\rho)
       -\sum_{j=1}^{d+1}P(A_j|\rho)\Big]} \\
     & = \sqrt{d(d^2-1)\left[dP(\rho)-1\right]},
  \end{aligned}
 \end{equation}
where the equality is due to $\sum_{j=1}^{d+1}P(A_j|\rho)=
1+P(\rho)$ when $d$ is a power of a prime \cite{purity}. By further
choosing $P(\rho)=1$, one can obtain a strongest state-independent
bound as
\begin{equation} \label{eq3a-3}
  \sum_{j=1}^{d+1}C_{l_1}^{A_j}(\rho)
         \leqslant (d-1)\sqrt{d(d+1)}
         \coloneqq C_{l_1}^m,
\end{equation}
and it reduces to that of Ref. \cite{NAQC} when $d=2$. So if one
considers the $l_1$ norm of coherence, $C_{l_1}^{na} (\rho_{AB})>
C_{l_1}^m$ is a signature of NAQC existing in the state $\rho_{AB}$.

\subsection{Relative entropy of coherence} \label{sec:3b}
We first prove a lemma concerning the relation between the von
Neumann entropy $S(\rho)$ and purity $P(\rho)$ for the general
$d$-dimensional state $\rho$. By denoting $\{\lambda_j\}$ for the
eigenvalues of $\rho$, $S(\rho)$ and $P(\rho)$ can be written
explicitly as
\begin{eqnarray} \label{eq3b-1}
  S(\rho)= -\sum_j \lambda_j \log_2 \lambda_j,~~
  P(\rho)= \sum_j \lambda_j ^2,
\end{eqnarray}
then by using the inequality $-\log_2 x\geqslant (1-x)/\ln 2$
($\forall x\geqslant 0$), one can show that
\begin{equation} \label{eq3b-2}
 \begin{aligned}
  S(\rho)+P(\rho) &\geqslant \sum_j \lambda_j \frac{1-\lambda_j}{\ln 2}+ \sum_j \lambda_j^2 \\
                  &= \frac{1}{\ln 2}\Big[ 1+(\ln 2 -1)\sum_j\lambda_j^2 \Big]
                  \geqslant 1.
 \end{aligned}
\end{equation}

By combining the above inequality with the complementarity relation
for the relative entropy of coherence \cite{comple1}, one can obtain
\begin{equation} \label{eq3b-3}
  \begin{aligned}
   \sum_{j=1}^{d+1}C_{re}^{A_j}(\rho)
     \leqslant & (d+1)[\log_2{d}+P(\rho)-1] \\
               & - \frac{(d-1)\log_2(d-1)}{d(d-2)}[dP(\rho)-1],
  \end{aligned}
\end{equation}
where $C_{re}^{A_j}(\rho)$ denotes the relative entropy of coherence
in the basis spanned by the eigenvectors of $A_j$. By further
choosing $P(\rho)=1$, one can obtain the strongest state-independent
upper bound as
\begin{equation} \label{eq3b-4}
 \sum_{j=1}^{d+1}C_{re}^{A_j}(\rho)
   \leqslant (d+1)\log_2{d}-\frac{(d-1)^2\log_2(d-1)}{d(d-2)}
   \coloneqq C_{re}^m,
\end{equation}
and for the special case $d=2$, we have $C_{re}^m=3-\log_2 e/2$, but
this bound can be further sharpened to $3H(1/2+\sqrt{3}/6)$
\cite{comple1}. So if one uses the relative entropy of coherence,
$C_{re}^{na}(\rho)> C_{re}^m$ captures the NAQC in $\rho_{AB}$.

\section{New framework of NAQC} \label{sec:4}
\begin{figure}
\centering
\resizebox{0.43 \textwidth}{!}{%
\includegraphics{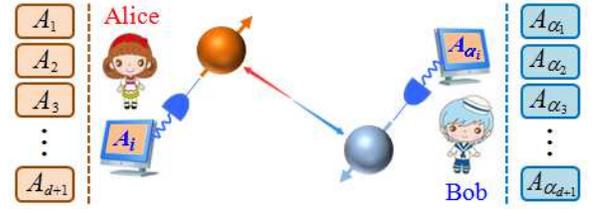}}
\caption{Schematic picture of our ``coherence steering game". There
is a one-to-one correspondence between Alice's measurement $A_i$ and
Bob's measurement basis $A_{\alpha_i}$ ($\forall i$), and
$\{\alpha_i\}$ is any possible permutations of the elements of
$\{i\}$.} \label{fig:1}
\end{figure}

In this section, we present a new framework for formulating the
NAQC. Different from that of Sec. \ref{sec:3} in which Bob's chosen
basis may be spanned by any $A_j$ of the set $\{A_j\}_{j\neq i}$
when Alice executed one round of the measurements and announced her
choice $A_i$ and outcomes $a\in\{1,\ldots,d\}$, in this new
framework Bob measures the coherence of his conditional states in
the preagreed basis spanned by the eigenvectors of $A_{\alpha_i}$,
where $\{\alpha_i\}$ is one of the possible permutations of the
elements of $\{i\}$. That is, there should be a one-to-one
correspondence between $i$ and $\alpha_i$ ($\forall i$). This is
illustrated in Fig. \ref{fig:1}.

Within this new framework, Bob's coherence averaged over all his
possible conditional states due to Alice's different measurements
can be written as
\begin{equation} \label{eq4-1}
 C^{na:\{\alpha_i\}}(\rho_{AB})= \sum_{i,a} p(a|A_i)
                C^{A_{\alpha_i}}(\rho_{B|A_i^a}),
\end{equation}
and it applies to any faithful coherence measure $C$. Moreover, for
the special case of $d=2$, $C^{na:\{\alpha_i\}}(\rho_{AB})$ can also
be obtained analytically (see Appendix \ref{sec:B}).

As for all the single-partite states $\rho$ of dimension $d$, we
have $\sum_i C^{A_{\alpha_i}}(\rho) \leqslant C^m$, the criterion
for achieving NAQC becomes $C^{na:\{\alpha_i\}}(\rho_{AB})>C^m$. It
holds for any possible $\{\alpha_i\}$, so one can optimize
$C^{na:\{\alpha_i\}}(\rho_{AB})$ over all the possible
$\{\alpha_i\}$ and define
\begin{equation} \label{eq4-2}
 \tilde{C}^{na}(\rho_{AB})=\max_{\{\alpha_i\}} C^{na:\{\alpha_i\}}(\rho_{AB}),
\end{equation}
then an optimized criterion is obtained as
\begin{equation} \label{eq4-3}
 \tilde{C}^{na}(\rho_{AB})> C^m.
\end{equation}

\begin{figure}
\centering
\resizebox{0.43 \textwidth}{!}{%
\includegraphics{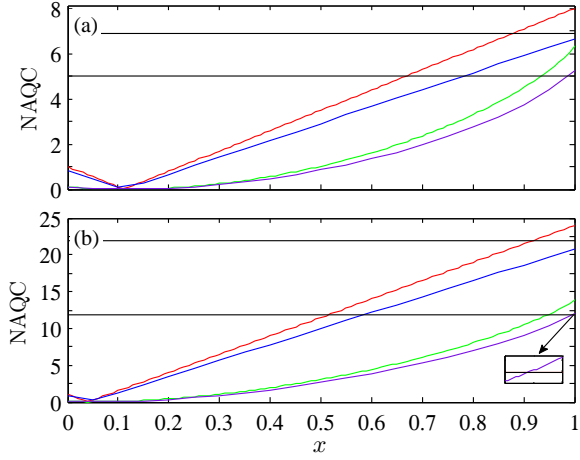}}
\caption{The NAQC for $\rho_I$ of Eq. \eqref{eq4-4} with $d=3$ (a)
and $5$ (b). For every plot, the red, blue, green, and purple (from
top to bottom) lines correspond to $\tilde{C}_{l_1}^{na} (\rho_I)$,
$C_{l_1}^{na} (\rho_I)$, $\tilde{C}_{re}^{na} (\rho_I)$, and
$C_{re}^{na} (\rho_I)$, while the two black horizontal lines
represent the corresponding bounds $C_{l_1}^{m}$ (top) and
$C_{re}^{m}$ (bottom), respectively.} \label{fig:2}
\end{figure}

This criterion could capture a wider regime of NAQC states than that
of Eq. \eqref{eq3-3} for certain cases. As an explicit example, we
consider the $(d\times d)$-dimensional isotropic state \cite{iso}
\begin{equation} \label{eq4-4}
 \rho_I=\frac{1-x}{d^2-1}I_{d^2}+\frac{d^2 x-1}{d^2-1}
                   |\Phi\rangle\langle \Phi|,~~ x\in[0,1],
\end{equation}
where $|\Phi\rangle= \sum_n |nn\rangle/\sqrt{d}$, and
$\{|n\rangle\}$ denotes the computational basis on $\mathbb{C}^d$.
By adopting the $l_1$ norm and the relative entropy as measures of
coherence, we calculated $\tilde{C}^{na}(\rho_I)$ and
$C^{na}(\rho_I)$. For $d=2$, we found that Eqs. \eqref{eq3-3} and
\eqref{eq4-3} capture the same region of $\rho_I$ that achieve NAQC.
For $d=3$ and 5, as can be seen from Fig. \ref{fig:2}, Eq.
\eqref{eq4-3} captures a wider region of $\rho_I$ with NAQC than
that of Eq. \eqref{eq3-3}. In particular, when one uses the $l_1$
norm of coherence, Eq. \eqref{eq3-3} cannot capture the NAQC in
$\rho_I$ at all.

For any separable state $\rho_\mathrm{sep}$, in a similar manner to
proving Eq. \eqref{eq3-5}, one can show that
\begin{equation} \label{eq4-5}
 C^{na:\{\alpha_i\}}(\rho_\mathrm{sep})
    \leqslant \sum_{k,i,a} q_k p_k(a|A_i) C^{A_{\alpha_i}}(\rho_B^k)
    \leqslant C^m,
\end{equation}
from which one can further obtain
\begin{equation} \label{eq4-6}
 \tilde{C}^{na} (\rho_\mathrm{sep})< C^m,
\end{equation}
thus Bob cannot achieve the NAQC for all the separable states even
in the new framework. This implies that all the $(d\times
d)$-dimensional states $\rho_{AB}$ which can achieve the NAQC at a
part of the system are quantum entangled.

Compared with the original framework, the new framework we
constructed also has the advantage of being easy to implement in
experiments. This is because after every round of Alice's
measurements $A_i$, Bob only needs to measure coherence of his
conditional states with respect to the eigenbasis of $A_{\alpha_i}$.
But for the original framework, the number of Bob's reference bases
is $d$ for every round of Alice's measurements.

Experimentally, the NAQC of $\rho_{AB}$ can be estimated by the
conditional tomography on \textit{B}, and this is easier than the
full state tomography. So Eq. \eqref{eq4-3} can also be used for
witnessing entanglement. That is, whenever we observed
$\tilde{C}^{na}(\rho_{AB})> C^m$, we are sure that $\rho_{AB}$ is
entangled. In particular, this criterion can witness different
regions of entanglement compared with other witnesses. As a
comparison, we consider the witnesses of tomographic estimate,
measurement estimate, and Fano estimate constructed via the entropic
uncertainty principle \cite{EUR1,EUR2,EUR3}. For $d>2$, both the
measurement estimate and Fano estimate cannot witness entanglement
in $\rho_I$, while Eq. \eqref{eq4-3} can always witness a finite
region of entanglement, although it is narrower than that witnessed
by the tomographic estimate (see Appendix \ref{sec:C}). For certain
states, Eq. \eqref{eq4-3} may witness entanglement that cannot be
witnessed by all the three uncertainty estimates, e.g., for $\rho_1$
considered in Ref. \cite{EUR2}
\begin{equation} \label{eq4-8}
 \rho_1=x|\Phi^+\rangle\langle\Phi^+|+(1-x)|\Psi^-\rangle\langle\Psi^-|,~~ x\in[0,1],
\end{equation}
where $|\Phi^+\rangle=(|00\rangle+|11\rangle)/\sqrt{2}$ and
$|\Psi^-\rangle=(|01\rangle-|10\rangle)/\sqrt{2}$, the regions of
entanglement witnessed by the three uncertainty estimates are
$x<0.110$ and $x>0.890$, and that witnessed by $\tilde{C}^{na}_{l_1}
> C_{l_1}^m$ is $x<0.138$ and $x>0.862$, see Appendix \ref{sec:C}.

Furthermore, as the NAQC may be enhanced by performing local unitary
operation $U_A\otimes U_B$ on $AB$ and the entanglement is a locally
unitary invariant, we are sure that $\rho_{AB}$ is entangled
provided
\begin{equation} \label{eq4-9}
 \max_{\{U_A\otimes U_B\}}\tilde{C}^{na}(U_A\otimes U_B \rho_{AB}
   U_A^\dag \otimes U_B^\dag) > C^m.
\end{equation}
Of course, one can also maximize $C^{na}(\rho_{AB})$ over all local
unitaries $\{U_A\otimes U_B\}$, but the witnessed region may be
different.

\section{Summary and discussion} \label{sec:5}
In summary, we have derived the criteria which capture the NAQC in a
bipartite state. We first generalized the framework of Mondal
\textit{et al.} \cite{NAQC} to $(d\times d)$-dimensional states with
$d$ being any power of a prime, then derived the explicit criteria
by considering the $l_1$ norm of coherence and the relative entropy
of coherence. We also presented a new framework for formulating NAQC
which can capture a wider regime of NAQC states than that of the
original one. Within both the two frameworks, we showed that any
state with NAQC is entangled, so one can recognize what the NAQC
captures as a kind of quantum correlation which is stronger than
entanglement, and the criteria can also be regarded as  as an
entanglement witness. We hope these results may lead to a better
understanding of the interrelation between coherence and quantum
correlations.

We remark that there are other faithful coherence measures
\cite{Plenio,Hu}. Their complementarity relations under mutually
unbiased bases and the criteria for achieving NAQC could be expected
in future research. Moreover, the present criteria apply to the case
of $d$ being a prime power, and they capture different sets of the
NAQC states when one uses different measures of coherence. A general
criterion which is applicable to bipartite states of arbitrary
dimension still remains to be explored, while seeking a
measure-independent criterion is also of practical significance as
the NAQC is hoped to serve as a resource for quantum information
processing.

\section*{ACKNOWLEDGMENTS}
This work was supported by NSFC (Grants No. 11675129 and No.
91536108), MOST (Grants No. 2016YFA0302104 and No. 2016YFA0300600),
New Star Project of Science and Technology of Shaanxi Province
(Grant No. 2016KJXX-27), CAS (Grants No. XDB01010000 and No.
XDB21030300), and New Star Team of XUPT.

\begin{appendix}
\section{The mutually unbiased bases} \label{sec:A}
\setcounter{equation}{0}
\renewcommand{\theequation}{A\arabic{equation}}
When the dimension $d$ of a system is a prime number, there are
$d+1$ mutually unbiased bases. By denoting $|\phi_m^l\rangle$ the
$m$th vector ($m=0,\ldots, d-1$) in the $l$th basis, we have
\cite{MUB1}
\begin{equation}\label{eqa-1}
 \begin{aligned}
  & |\phi_m^0\rangle= \sum_{n=0}^{d-1}\delta_{mn} |n\rangle,~~
    |\phi_m^d\rangle= \frac{1}{\sqrt{d}}\sum_{n=0}^{d-1} e^{i\frac{2\pi}{d}mn}|n\rangle,\\
  & |\phi_m^l\rangle= \frac{1}{\sqrt{d}}\sum_{n=0}^{d-1} e^{i\frac{2\pi}{d}l(m+n)^2}|n\rangle,
                      ~~ \textrm{for} ~~ l=1,\ldots,d-1,
 \end{aligned}
\end{equation}
while for $d$ being a prime power, there are also $d+1$ mutually
unbiased bases which have been constructed in Ref. \cite{MUB2}.

Based on these mutually unbiased bases, one can obtain the ensemble
of Bob's conditional states as $\{p(a|A_i),\rho_{B|A_i^a}\}$, where
the postmeasurement state of \textit{B} is
\begin{equation}\label{eqa-2}
 \rho_{B|A_i^a}=\frac{\langle\phi_a^i| \rho_{AB} |\phi_a^i\rangle}{p(a|A_i)},
\end{equation}
and $p(a|A_i)$ is the probability for Alice's outcome $a$ when she
measures $A_i$. It can be written as
\begin{equation}\label{eqa-3}
 p(a|A_i)= \mathrm{tr}\big(\langle\phi_a^i| \rho_{AB} |\phi_a^i\rangle\big).
\end{equation}

If the bipartite state is separable, i.e., $\rho_\mathrm{sep}=\sum_k
q_k \rho_A^k \otimes \rho_B^k$, one can obtain
\begin{equation} \label{eqa-4}
 \rho_{B|A_i^a}= \frac{\sum_k q_k \langle\phi_a^i| \rho_A^k |\phi_a^i\rangle \rho_B^k}
                   {\sum_k q_k \mathrm{tr}\big(\langle\phi_a^i| \rho_A^k\otimes\rho_B^k |\phi_a^i\rangle \big)}
                 = \frac{\sum_k q_k p_k(a|A_i) \rho_B^k}
                   {\sum_k q_k p_k (a|A_i)}.
\end{equation}
where we have defined $p_k (a|A_i)=\langle\phi_a^i| \rho_A^k
|\phi_a^i\rangle$.

\section{Solution of Eq. (15) for two-qubit states} \label{sec:B}
\setcounter{equation}{0}
\renewcommand{\theequation}{B\arabic{equation}}
By denoting $\vec{r}$ and $\vec{s}$ the local Bloch vectors, and
$\vec{\sigma}$ the vector of Pauli operators,  one can decompose a
general two-qubit state $\rho_{AB}$ as follows
\begin{equation} \label{eqb-1}
 \rho_{AB}= \frac{1}{4}\Big(I_4+\vec{r}\cdot\vec{\sigma}\otimes I_2
       +I_2\otimes\vec{s}\cdot\vec{\sigma}
       +\sum_{i,j} t_{ij}\sigma_i\otimes\sigma_j\Big),
\end{equation}
where $r_i=\mathrm{tr}\rho_{AB}(\sigma_i\otimes I_2)$,
$s_i=\mathrm{tr}\rho_{AB}(I_2\otimes\sigma_i )$, and
$t_{ij}=\mathrm{tr}\rho_{AB}(\sigma_i\otimes \sigma_j)$
($i,j=1,2,3$).

Based on this decomposition, the probability of Alice's outcome $a$
when she measures $A_i$ can be obtained as \cite{NAQC}
\begin{equation} \label{eqb-2}
 p(a|A_i)= \frac{1+(-1)^a r_i}{2}.
\end{equation}
then the $l_1$ norm of coherence for the conditional state
$\rho_{B|A_i^a}$ is given by
\begin{equation} \label{eqb-3}
 C^{A_{\alpha_i}}_{l_1}(\rho_{B|A_i^a})=\frac{\sqrt{\sum_{j\neq \alpha_i}
                                    [s_j+(-1)^a t_{ij}]^2}}{1+(-1)^a r_i},
\end{equation}
and the relative entropy of coherence for $\rho_{B|A_i^a}$ is given
by
\begin{equation} \label{eqb-4}
 C^{A_{\alpha_i}}_{re}(\rho_{B|A_i^a})= H(\beta_{i a})-H(\lambda_{i a}),
\end{equation}
where $H(\cdot)$ is the binary Shannon entropy function, and
\begin{equation} \label{eqb-5}
 \begin{aligned}
   & \beta_{i a}= \frac{1}{2}+ \frac{s_{\alpha_i}+(-1)^a t_{i \alpha_i}}
                         {2[1+(-1)^a r_i]}, \\
   & \lambda_{i a}= \frac{1}{2}+ \frac{\sqrt{\sum_j [s_j+(-1)^a t_{i j}]^2}}
                   {2[1+(-1)^a r_i]}.
 \end{aligned}
\end{equation}

Finally, by substituting Eqs. \eqref{eqb-3} or \eqref{eqb-4} into
Eq. \eqref{eq4-1}, we obtain $C^{na:\{\alpha_i\}}(\rho_{AB})$ for a
given $\{\alpha_i\}$, and by optimizing over all possible
$\{\alpha_i\}$, one can further obtain $\tilde{C}^{na}(\rho_{AB})$.

\section{Application of the criteria for witnessing entanglement} \label{sec:C}
\setcounter{equation}{0}
\renewcommand{\theequation}{C\arabic{equation}}

For two chosen observables $R$ and $S$ performed on $A$ of
$\rho_{AB}$, the entropic uncertainty relation reads \cite{EUR1}
\begin{equation} \label{eqc-1}
 H(R|B)+H(S|B)\geqslant -\log_2 c +H(A|B),
\end{equation}
where $H(A|B)$ is the conditional entropy of $\rho_{AB}$, $H(X|B)$
($X=\{R,S\}$) is the conditional entropy of the postmeasurement
state $\rho_{XB}$, and $c=\max_{k,l}|\langle \psi_k
|\phi_l\rangle|^2$, with $\{|\psi_k\rangle\}$ and
$\{|\phi_l\rangle\}$ being the eigenvectors of $R$ and $S$,
respectively.

Based on the above equation, Berta \textit{et al.} proposed to
witness entanglement via the following inequality
\begin{equation} \label{eqc-2}
   E_\alpha< -\log_2 c~~ (\alpha=\mathrm{T,~ M,~ or~ F}), \\
\end{equation}
where $E_\mathrm{T}=H(R|B)+H(S|B)$, $E_\mathrm{M}= H(R|R)+H(S|S)$,
and $E_\mathrm{F}=H(p_R)+H(p_S)+(p_R+p_S) \log_2(d-1)$ represent the
tomographic estimate, measurement estimate, and Fano estimate of the
uncertainty, respectively. Moreover, $p_X$ is the probability that
the outcomes of $X$ on $A$ and $X$ on $B$ are different. Whenever
Eq. \eqref{eqc-2} is fulfilled, one is sure that $\rho_{AB}$ is
entangled. In the following, we fix $R=A_1=I_d$ and $S=A_i$ ($i\neq
1$), where the meaning of $A_i$ is the same as that in the main
text.

For state $\rho_I$ of Eq. \eqref{eq4-4} with $d=2$, we have
$E_\alpha=2H(\eta)$ ($\forall \alpha$), $\tilde{C}_{l_1}^{na}
(\rho_I)=|4x-1|$, and $\tilde{C}_{re}^{na} (\rho_I)=3-3H(\eta)$,
where $\eta=(1+2x)/3$. Then one can obtain that the entanglement
region witnessed by $E_\alpha< -\log_2 c$ ($x>0.835$) is wider than
that witnessed by $\tilde{C}^{na}_{l_1}> C^m_{l_1}$ ($x>0.862$) and
$\tilde{C}^{na}_{re}> C^m_{re}$ ($x>0.935$). For $d\geqslant 3$, the
numerical results reveal that the entanglement region witnessed by
$E_\mathrm{T}< -\log_2 c$ is still wider than that witnessed by Eq.
\eqref{eq4-3}, but $E_\mathrm{M,E}< -\log_2 c$ cannot witness
entanglement in $\rho_I$.

For $\rho_1$ of Eq. \eqref{eq4-8}, we have $E_\alpha=2H(x)$
($\forall \alpha$), $\tilde{C}_{l_1}^{na}(\rho_1)=1+|4x-2|$, and
$\tilde{C}_{re}^{na}(\rho_1)=3-2H(x)$, from which one can obtain
that the entanglement regions witnessed by $\tilde{C}^{na}_{l_1}>
C^m_{l_1}$ ($x<0.138$ and $x>0.862$) are wider than that witnessed
by $E_\alpha< -\log_2 c$ ($x<0.110$ and $x>0.890$), while the
regions witnessed by $\tilde{C}^{na}_{re}> C^m_{re}$ ($x<0.075$ and
$x>0.925$) are narrower than that witnessed by $E_\alpha< -\log_2
c$.

\end{appendix}

%
\newcommand{\PRL}{Phys. Rev. Lett. }
\newcommand{\RMP}{Rev. Mod. Phys. }
\newcommand{\PRA}{Phys. Rev. A }
\newcommand{\PRB}{Phys. Rev. B }
\newcommand{\PRE}{Phys. Rev. E }
\newcommand{\PRX}{Phys. Rev. X }
\newcommand{\NJP}{New J. Phys. }
\newcommand{\JPA}{J. Phys. A }
\newcommand{\JPB}{J. Phys. B }
\newcommand{\PLA}{Phys. Lett. A }
\newcommand{\NP}{Nat. Phys. }
\newcommand{\NC}{Nat. Commun. }
\newcommand{\SR}{Sci. Rep. }
\newcommand{\EPJD}{Eur. Phys. J. D }
\newcommand{\QIP}{Quantum Inf. Process. }
\newcommand{\QIC}{Quantum Inf. Comput. }
\newcommand{\AoP}{Ann. Phys. }
\newcommand{\PR}{Phys. Rep. }
%

%

\end{document}